\documentclass[10pt,twocolumn,a4paper,aps,pra]{revtex4-1}
\usepackage[utf8]{inputenc}
\usepackage{graphicx}
\usepackage{color}
\usepackage{subscript}
\usepackage{amsmath,amssymb}
\usepackage{lineno}
\begin{document}

%%%%%%%%%%%%%%%%%%%%%%%%%%%%%%%%%%%%%%%%%%%%%%%%%%%%%%%
%%%%%%%%%%%%%%%%%%%%%%%%%%%%%%%%%%%%%%%%%%%%%%%%%%%%%%%
%%%%%%%%%%%%%%%%%%%%%%%%%%%%%%%%%%%%%%%%%%%%%%%%%%%%%%%

\title{Particle-in-Cell/Test-Particle Simulations of Technological Plasmas: Sputtering Transport in Capacitive Radio Frequency Discharges}

\author{Jan Trieschmann}\author{Frederik Schmidt}\author{Thomas Mussenbrock}

\affiliation{Ruhr University Bochum, Department of Electrical Engineering and Information Science, 44780 Bochum, Germany}

\date{\today}

\begin{abstract}
The paper provides a tutorial to the conceptual layout of a self-consistently coupled Particle-In-Cell/Test-Particle model for the kinetic simulation of sputtering transport in capacitively coupled plasmas at low gas pressures. It explains when a kinetic approach is actually needed and which numerical concepts allow for the inherent nonequilibrium behavior of the charged and neutral particles. At the example of a generic sputtering discharge both the fundamentals of the applied Monte Carlo methods as well as the conceptual details in the context of the sputtering scenario are elaborated on. Finally, two in the context of sputtering transport simulations often exploited assumptions, namely on the energy distribution of impinging ions as well as on the test particle approach, are validated for the proposed example discharge.
\end{abstract}

\maketitle

%%%%%%%%%%%%%%%%%%%%%%%%%%%%%%%%%%%%%%%%%%%%%%%%%%%%%%%
%%%%%%%%%%%%%%%%%%%%%%%%%%%%%%%%%%%%%%%%%%%%%%%%%%%%%%%
%%%%%%%%%%%%%%%%%%%%%%%%%%%%%%%%%%%%%%%%%%%%%%%%%%%%%%%

\section{\label{sec:introduction} Introduction}

Within many technological process chains, low temperature plasmas are indispensably utilized. One of the most important feature of these plasmas (often referred to as technological plasma) is their inherent nonequilibrium property. The individual characteristic kinetic energies of the constituents of the plasma (electrons, ions, radicals, etc.) are diversely spread from meV for neutral background particles over eV for electrons and energetic neutrals (e.g., sputtered particles) up to keV in case of ions which have undergone acceleration by the electric field, which is present in the sheath at all material boundaries of the plasma. Of striking importance in this respect is that electromagnetic energy coupled into the plasma by external sources is predominantly coupled into the electrons. Due to their relatively low mass compared to ions and their corresponding small inertia, electrons respond to an external excitation almost instantaneously. In contrast, ions typically react on a much slower timescale and, in fact, oftentimes only on time averaged quantities. This aspect is of major concern regarding the operational principle of capacitively coupled plasmas (CCPs) driven by high-frequency electric fields. However, this shall not be elaborated in this work. Great reference is provided elsewhere.\cite{lieberman_principles_2005,chabert_physics_2011}

While the understanding of plasma physics based on experiments does legitimate in its own right (as \emph{all} physical phenomena are incorporated by nature), they often provide only indirect and/or limited insight due to limited access to the physical quantities of interest. Moreover, in particular in the context of industrial manufacturing the prevention of any kind of perturbation (e.g., due to applied diagnostic apparatuses) is a clear premise. Simulation approaches, however, offer powerful add-ons or even alternatives. At the expense of limited physical complexity, the most intrinsic physical processes are actually available for analysis. As such, theoretical models can be utilized for both prediction and optimization of relevant plasma processes. It is imperative to acknowledge that various conceptually different simulation methods are principally applicable in order to describe technological plasmas. Yet, not all methods are suitable to describe all physical phenomena on each length and time scale of interest.

The purpose of this paper is twofold: i) As a tutorial paper it provides a conceptual layout of a self-consistently coupled particle-in-cell/test-particle model which treats all species of the plasma in a kinetic framework and ii) as a research paper it reports on and discusses new results of the physics of sputtering transport in radio-frequency (rf) driven capacitive discharges at low gas pressures. The paper is organized as follows: Firstly, it is explained when a kinetic approach is needed and which numerical concepts actually allow for a description of the nonequilibrium behavior of the particles. Secondly, using the example of a sputtering discharge the fundamentals of Monte Carlo methods as well as the conceptual details in the context of the specified sputtering deposition scenario are elaborated on. Calculation results of the given exemplary case are presented and final conclusions are drawn.

%%%%%%%%%%%%%%%%%%%%%%%%%%%%%%%%%%%%%%%%%%%%%%%%%%%%%%%
%%%%%%%%%%%%%%%%%%%%%%%%%%%%%%%%%%%%%%%%%%%%%%%%%%%%%%%
%%%%%%%%%%%%%%%%%%%%%%%%%%%%%%%%%%%%%%%%%%%%%%%%%%%%%%%

\section{\label{sec:approaches} Simulation Concepts}

It is important to accept that not all simulation methods are applicable to the various kinds of plasmas and physical situations. Before formulating a model a careful analysis with respect to time and length scales is necessary. Based on a scale analysis one is able to find arguments for and against the choice of a particular model. For instance, for the energy relaxation length is crucial. It characterizes the length scale of Maxwellization.\cite{kortshagen_electron_2002} 

In low temperature plasmas collisions between charged particles are relatively rare so that collisions between charged particles and neutral particles dominate the dynamics. Ions and neutrals are of comparable mass so that elastic ion-neutral collisions significant energy transfer. The energy relaxation length for ions is therefore approximately equal to the mean free path for ion-neutral collisions. The situation is different for electrons. Due to the large mass difference between electrons and neutrals, only a small portion of energy is transferred by elastic collisions. That means that a large number of collisions are needed for Maxwellization of electrons. The energy relaxation length of electrons which only undergo elastic collisions with neutrals is approx. $\lambda_\epsilon\approx \lambda_{el,e}(m_g/m_e)^{1/2}$ where $\lambda_{el,e}$ is the mean free path for elastic electron-neutral collisions, while $m_g$ and $m_e$ are the mass of a neutral particle and the mass of an electrons, respectively. When inelastic collisions of electrons and neutrals are significant, the energy relaxation length will decrease. For larger average electron energies the contribution of inelastic collisions may dominate. When inelastic processes which include ionization, electron attachment, and electronic excitation int he case of atomic gases and additionally rotational excitation, vibrational excitation, and various dissociation processes in the case of molecular gases the electron energy relaxation length is approximately equal to the effective mean free path for all inelastic processes. The mean free path for each inelastic process depends of course on the details of the cross section for the process as well as on the electron energy distribution function. 

\begin{figure}[t!]
\includegraphics[width=8cm]{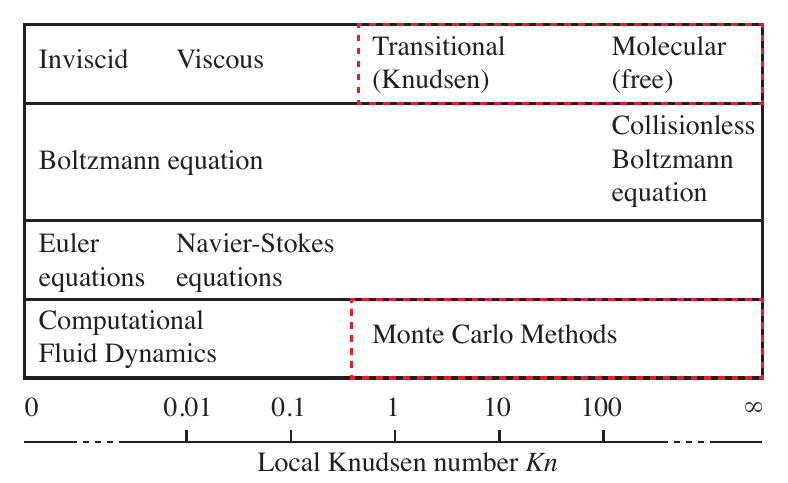}
\caption{Illustration of the respective physical regimes and the corresponding theoretical approaches in the form of a mathematical description as well as computer simulations. The figure is adopted from \cite{bird_molecular_1994}.}
\label{fig:knudsen_bird}
\end{figure}

For neutrals the energy relaxation length is in fact the mean free path. If the mean free path $\lambda$ is larger than the typical dimension of the discharge $L$ the characteristic Knudsen number $\text{Kn}=\lambda/L$ is larger than unity. Under these circumstances, a kinetic approach for the heavy particles is needed as well. The various modeling approaches and the respective Knudsen numbers are depicted in figure~\ref{fig:knudsen_bird}. (In his fundamental work on the direct simulation Monte Carlo method, Bird provides a very illustrative schematic of the physical regimes and conceptual models appropriate.\cite{bird_molecular_1994})

For the topic of this paper in particular the transitional flow regime and the free molecular flow regime and, thus, the bottom right part of figure~\ref{fig:knudsen_bird} are of importance. That is, when the collisional interaction is small but not negligible. In this respect, collisions can be particle interactions among each other, or as a particle trace within an abundant background gas. Collisions can also be interactions with the walls. In the transitional regime particle interactions can to good accuracy be assumed as binary (i.e., two particles directly interact). In contrast, for free molecular flow the trace particles merely interact with a background and can as such be treated as physically independent particles throughout the simulation. Despite very low gas pressures, in the situations encountered in the physics of low temperature plasmas, the problems not only involve neutral but also charged species. Hence, the constituents can be grouped into: (i) charged species (electrons and ions) and (ii) neutral species (thermal and fast neutrals, excited species, or sputtered neutral particles). Both groups can be described by means of Lagrangian computer models following the Monte Carlo approach.

\section{\label{sec:setup} Simulation Scenario}

Using the example of a capacitively coupled sputtering discharge driven by a radio-frequency power supply, all relevant components required for a consistent theoretical description shall be briefly discussed. This scenario is particularly chosen as it requests a number of peculiar aspects to be taken into account, which at the same time can be instructively decomposed individually. Moreover, the required background in plasma physics is intentionally kept to a most fundamental level of knowledge.

\begin{figure}[t!]
\includegraphics[width=8cm]{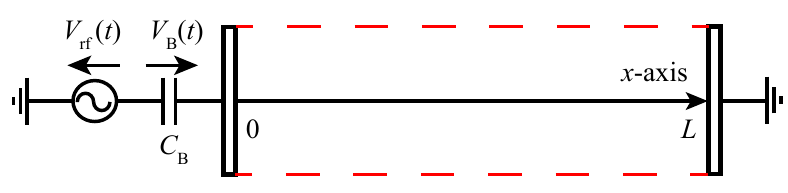}
\caption{One dimensional discharge setup consisting of two electrodes (driven: left, grounded: right) connected to an rf voltage supply $V_\textrm{rf}$ via a blocking capacitor $C_\textrm{B}$. The red corridor illustrates geometric invariance into lateral dimensions.}
\label{fig:setup}
\end{figure}

\paragraph{Discharge Setup.}
As depicted in figure~\ref{fig:setup}, the discharge consists of a one dimensional arrangement of two opposing plane, ideal electrodes of infinite surface area. These electrodes are separated by a gap of $L = 7.5$~cm. The left electrode is made of aluminum for the purpose of sputtering. It is referred to as the \emph{target}. The right electrode is made of stainless steel and is referred to as the \emph{substrate}. In the course of this paper, the discharge will be viewed under steady-state conditions. It is therefore important to acknowledge that during operation the stainless steel surface of the substrate electrode is in fact irrelevant, as its surface properties are governed by a layer of previously deposited aluminum covering the surface. For the purpose of sputtering, the discharge is operated with argon gas at a constant pressure $p=0.5$~Pa, assuming a background gas temperature of $T=650$~K.\cite{bienholz_multiple_2013,trieschmann_transport_2015} A gas density of $n_\textrm{Ar} \approx 5.57 \times 10^{13} \textrm{~cm}^{-3}$ consequently dominates the collisional sputtered particle transport of concern.

To drive the discharge an ideal voltage source $V_\textrm{rf}(t)$ is applied at the left electrode (connected via a blocking capacitor with $C_\textrm{B} = 1~\mu\textrm{F}$). A radio-frequency voltage waveform $V_\textrm{rf}(t) = V_0 \left[ \cos\left( \omega t + \Delta \phi \right) + \cos\left( 2 \omega t \right) \right]$ is applied with the angular frequency $\omega = 2 \pi f$, the driving frequency $f=13.56$~MHz, and a voltage amplitude $V_0 = 300$~V -- that is, a fundamental frequency and its harmonic. Utilizing the electrical asymmetry effect (EAE), this geometrically symmetric discharge can be made electrically asymmetric with a choice of the relative phase between the two harmonics of $\Delta \phi = 0$.\cite{heil_possibility_2008,czarnetzki_electrical_2009} The impact of which will become more clear in the course of this paper. The right electrode is held at ground potential.

\paragraph{Operational Regime.}
For what follows a principle understanding of the internal kinetics of the discharge is important:\cite{lieberman_principles_2005} (i) Within the CCP environment, the plasma is operated in a continuous rf mode. A periodic steady-state is strictly expected. The discharge is stationary but not static: During an rf cycle electrons are accelerated back and forth within the discharge gap. Due to their large inertia ions are almost unaffected by the transient field, but are influenced by the time averaged electric field. (ii) With a plasma density of $n_\textrm{e} \approx 4 \times 10^{9} \textrm{cm}^{-3}$ the degree of ionization is below $10^{-4}$ and, consequently, the background gas density can be safely assumed to be unaffected by the plasma (i.e., gas rarefaction due to ionization can be ignored \cite{raadu_ionization_2011}). (iii) An immanent consequence is a comparable density of sputtered aluminum: The sputtered particle flux is governed by the flux and energy distribution of ions impinging the surfaces and the corresponding sputtering yield (roughly $Y\approx 1$). The simplest assumption is the one of mono-energetic ions, which strike the surface perpendicularly. The flux of impinging argon ions and the flux of sputtered aluminum are, therefore, of comparable order of magnitude. Resulting, the influence of the sputtered particles on the background gas density in the form of a sputtering wind can be neglected as well.\cite{hoffman_sputtering_1985} The sputtered aluminum is merely a trace within the abundant argon gas.

\begin{figure}[t!]
\includegraphics[width=8cm]{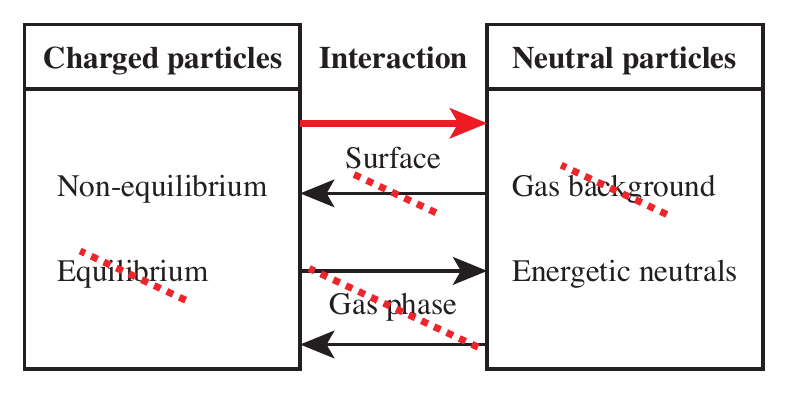}
\caption{Decomposition of a generic plasma discharge into only the relevant components and links for the depicted situation. The full coupling between the charged particle and the neutral particle dynamics is reduced to a unidirectional link via the sputtering surface interaction (red bold arrow).}
\label{fig:interactions}
\end{figure}

From a modeling point of view, the charged and neutral particle contributions can be understood as building blocks. For these, the above mentioned aspects do have imperative implications -- in terms of the individual modules as well as their interaction. The individual contributions can be interfaced solely through a reduced set of interactions at the surface and within the gas phase. This is schematically depicted in figure~\ref{fig:interactions}, where the full complexity of a generic plasma model is reduced to only one relevant link for the given discharge. (i) Ions impinging the surface may sputter surface atoms of the bulk material, which are the primary source of energetic neutral particles. In turn, neutral particles (particularly excited species) may generate secondary electrons when approaching the surface. This, however, adds only a small contribution to the interactions which charged particles experience at the surfaces anyhow. Consequently, the bidirectional link reduces to a unidirectional input from the charged particle to the neutral particle model. (ii) In the gas phase, from the above mentioned considerations, a-priori any dynamical effects within the background gas such as gas rarefaction can be neglected. Moreover, the sputtered particle concentration is so low as to render the amount of metal ions formed via electron-impact ionization insignificant (despite a lower ionization energy threshold). The corresponding bidirectional link fully resolves in favor of the straightforward assumption of a constant, stationary argon gas background for both charged and neutral sputtered particles. (Resonant charge-exchange collisions of argon ions within the given argon background are taken into account nevertheless.) The coupling between the individual modules diminishes greatly as feedback onto the charged particles may be ignored completely.

In terms of modeling, two separate modules may be implemented, one dealing with a dynamical description of the charged species and the other with the sputtered neutrals. The principle interfacing stems from the ion flux incident onto the adjacent surfaces, thus serving as input for the neutral particle module. For the sake of generality it should be noted that, by proper interfacing of the full linkage, a consistent description may be recovered if required. A self-consistent model would then need to account for a transient neutral gas density within the plasma module and a closely coupled loop were to resemble.

\paragraph{Hypotheses.}
Within the preceding analysis, two important hypotheses have been postulated. (A) Firstly, it has been proposed that ions accelerated by the strong electric field in the sheaths, hit the target nearly mono-energetically and predominantly in surface-normal direction. This aspect concerns mostly the input to the test particle model utilized for the sputtered particle transport simulation.\cite{trieschmann_transport_2015}. (B) Secondly, whenever test particle simulations are carried out, the fundamental assumption is that the test particle species are only a trace within an abundant background. In consequence, the test particles are influenced by the abundant species, but not vice versa. The link is realized unidirectional and the feedback of the trace particles onto the background is ignored. Both of which hypotheses are in general highly relevant in the context of Monte Carlo sputtering plasma simulations. The benefit of a coupled particle-in-cell/test-particle model lies within the fact that both assumptions may (at least up to the fundamental arguments) be validated. This topic is to be discussed in the following sections, where related results are presented.

\section{\label{sec:mcsimulations} Kinetic Particle Simulation of Sputtering Plasmas}

In terms of a theoretical model, the previous considerations raise the question of how to constitute the individual modules. At the low pressures of interest, the evaluation of the dynamics of low temperature plasmas commonly requires a kinetic description of the many-particle system.\cite{kaganovich_stochastic_1996} Closed analytic formulations of the underlying complicated multiphysics/multiscale problems are not possible and even numerical solutions present an expensive computational task. In this paper, answers are sought by means of Monte Carlo computer simulations, which have proven to be readily applicable for tracing the system evolution over discrete times.\cite{metropolis_monte_1949,alexander_direct_1997,gould_introduction_2007,kersch_transport_2011,garcia_numerical_2015} 

Depending on the problem and physics of interest a simple 1-dimensional model might be a sufficient approximation. If 2 or 3 dimensions are to be taken into account, the simulation time increases. Within a $d$ dimensional simulation space the number of computational operations to be performed scales to the power $d$. Usually, multi-dimensional simulations are highly parallelized to cope with the computational demands. Within this paragraph, the common fundamental aspects of both the charged and the neutral particle module are reviewed. Subsequently, the decisive specifics of the two are of concern in the respective subsections.

The overall idea of a Lagrangian particle (flow) picture is as follows:\cite{birdsall_plasma_1985,hockney_computer_1988,birdsall_particle--cell_1991} A given set of super-particles -- each of which represents a collection of physical particles -- is randomly distributed in phase space. Correspondingly, these super-particles represent samples of the distribution function of the species.\cite{metropolis_monte_1949,feller_introduction_1971} From the ergodic principle the former defines the latter and vice versa.\cite{cercignani_boltzmann_1988} Most simply, the super-particles are initially uniformly distributed in configuration space with position vectors $\vec{r}_k$ and Maxwellian distributed in velocity space with ``position'' vectors $\vec{v}_k$. Clearly, $k$ indicates an individual super-particle out of the full set of $K$ super-particles. From the statistical phase space sampling of the initial particle distribution, the term \emph{Monte Carlo} emerged.

Various methods (e.g., Box-Muller or Marsaglia polar method) have been proposed for the sampling of a thermal distribution from random uniformly distributed samples $R_\textrm{f} \in \left[ 0, 1 \right]$, which can be readily applied for the latter purpose.\cite{box_note_1958,marsaglia_convenient_1964} After initialization, the super-particles are traced in configuration space according to (non-relativistic) Newton's laws
\begin{align}
	\frac{\partial \vec{r}_k}{\partial t} &= \vec{v}_k, \label{eq:newtonfirstlaw}\\
   	\frac{\partial \vec{v}_k}{\partial t} &= \vec{a}_k. \label{eq:newtonsecondlaw}
\end{align}
Here, $\vec{a}_k = \vec{F}_k / m_k$ is the acceleration of the super-particle due to forces of any kind. For a gravitational force $\vec{F}_k = m_k \, \vec{g}$ the mass dependence forthrightly cancels. Under the influence of an electromagnetic field the force is given by the Lorentz force $\vec{F}_k = Q_k( \vec{E} + \vec{v}_k \times \vec{B})$. The acceleration acting on a physical particle scales with its charge $Q_k$ and inversely with its mass $M_k$. Physical particles with $Q_k/M_k$ therefore experience the same acceleration as the representative super-particles with $q_k/m_k$, given that both are based on the same weight $w=q_k/Q_k$. (Here $q_k= w \, Q_k$ is the charge of the super-particle, $m_k = w \, M_k$ its mass). In plasma simulations (accept for dusty plasmas, e.g., \cite{shukla_introduction_2001}), the force taken into account is prevalently reduced to the Lorentz force. Only charged species experience acceleration, neutral species are unaffected as gravitational forces are neglected.

By nature of the computational operations to be performed, a discrete time stepping is required. Within the frame of an explicit evaluation scheme, a so-called leap frog algorithm may be used.\cite{birdsall_clouds--clouds_1969,birdsall_plasma_1985} Particle positions are evaluated at multiples of the time interval $\Delta t$, while corresponding velocities are evaluated at times shifted by $\Delta t / 2$. Equations \eqref{eq:newtonfirstlaw} and \eqref{eq:newtonsecondlaw} may thus be decoupled following a second order central finite difference scheme (with $l$ indicating the discrete time of evaluation),
\begin{align}
	\vec{r}_k^{\,(l+1)} &= \vec{r}_k^{\,(l)} + \Delta t \, \vec{v}_k^{\,(l+1/2)} \label{eq:pushinr} \\
	\vec{v}_k^{\,(l+1/2)} &= \vec{v}_k^{\,(l-1/2)} + \Delta t \, \vec{a}_k^{\,(l)} \label{eq:pushinv}.
\end{align}
In the presence of a magnetic field, the right hand side of \eqref{eq:pushinv} requires additional considerations, for example following a procedure proposed by Boris.\cite{boris_relativistic_1970} Alternative schemes may as well be applied to solve the equations of motion \eqref{eq:newtonfirstlaw} and \eqref{eq:newtonsecondlaw}, for instance implicit methods relaxing certain stability criteria.\cite{wirz_discharge_2005,langdon_direct_1983} Essentially, the ensemble of super-particles is evolved in time and space subject to internal and external force fields.

To describe the system evolution appropriately, collisions play a crucial role. It is important to understand the dynamics of the system from a conceptual standpoint. At a given instant of time, that dynamical state of each super-particle, once in a while also referred to as simulator, is fully specified by its position and velocity vector and accordingly located in phase space. During the free flight of duration $\tau$, the collision time, the trajectory of every super-particle may be traced. Within every iteration cycle super-particles are first traced in phase space, then according to the probability and dynamics of a collision encounter is evaluated.

Following the assumption of a Markov chain, the occasion of an event solely depends on the present state and not the past. The sequence of statistically independent events (i.e., collision encounters) is then based on an exponential probability density $p_\textrm{c}(\tau) = \exp(-\tau/\tau_\textrm{c})$.\cite{feller_introduction_1971} From the collision probability 
\begin{align}
	P_\mathrm{c}(\tau \leq \tau_\textrm{c}) = 1 - \exp(-\tau/\tau_\mathrm{c}), \label{eq:collisionprobability1}
\end{align}
the collision time $\tau = -\tau_\textrm{c} \ln \left( R_\textrm{f} \right)$ can be evaluated using $R_\textrm{f}$, a sample from a uniformly random distribution. $\tau$ is, consequently, based on the expectation collision time $\tau_\textrm{c}$ (i.e., the expectation value of the probability distribution). The expectation collision time $\tau_\textrm{c} = \tau_\textrm{c}(v_\textrm{r})$ is specific to the relative speed $v_\textrm{r} = \left| \vec{v}_k - \vec{v}_\textrm{bg} \right|$ of particle $k$, which is inherently known, and its background collision partner (indicated by ``bg''). At any rate, it is not yet an ensemble mean. In case of a thermal neutral gas background, the velocity of the background particles $\vec{v}_\textrm{bg}$ may be again sampled from a thermal distribution. The expectation collision time is obtained from $\tau_\textrm{c} = 1 / \left( n_\mathrm{bg} \, \sigma(v_\mathrm{r}) \, v_\mathrm{r} \right)$, which is a function of the background density $n_\mathrm{bg}$, the velocity (or energy) dependent cross-section $\sigma(v_\mathrm{r})$ and the relative speed $v_\textrm{r}$. It is the inverse of the collision frequency $\nu_\mathrm{c} = 1 / \tau_\textrm{c}$. Concerning the ensemble average, this is consistent with the expectation value for an individual particle to collide with the ensemble of background particles obtained from the Boltzmann collision operator, given by $\nu_\textrm{c,mean} = n_\textrm{bg} \, \overline{\sigma(v_\textrm{r}) v_\textrm{r}}$. The overbar indicates the ensemble average over the velocity distribution of the particles within a control volume. Under acceleration free motion, the individual particle may be traced for a randomly selected time of free flight $\tau$, terminated by a collision. The corresponding distance of free flight is $\lambda = v_k \, \tau$.

In the presence of an accelerating force -- in particular, if it is spatially inhomogeneous -- the super-particle tracing procedure has to account for this peculiarity. By employing the staggered grid approach and a constant time step $\Delta t$, acceleration may be consistently taken into account. The collision probability is checked every $\Delta t$, whereas a collision is evaluated only when $R_\textrm{f} \leq P_\textrm{c}(\Delta t \leq \tau_c)$. The upper bound for the choice of $\Delta t$ is typically constrained by the scale of the spatial inhomogeneity $L$ of the force field. By requiring $\lambda \ll L$ often  $\Delta t \ll \tau_\textrm{c}$ is inferred. This is, in fact, a manifestation of displeasing inefficiency.

Due to its far reaching implications, the prevalent assumption of taking into account only collisions between plasma and respectively energetic neutral species with a thermal neutral gas background may need to be reconsidered for different situations (e.g., ionized physical vapor deposition). The individual aspects specific to the plasma and neutral species are left for later consideration in subsections \ref{sec:pic} and \ref{sec:tpm}, where also the interactions with the walls are considered. To conclude, the described iterative procedure is evolved until a steady-state situation is reached. Finally, the particle distributions in configuration and velocity space can be obtained by averaging over an adequate period of time. If a periodic steady-state is of concern, time-resolved averaging may be performed accordingly.

\subsection{\label{sec:pic} Particle-In-Cell Scheme}

\begin{figure}[t!]
\includegraphics[width=8cm]{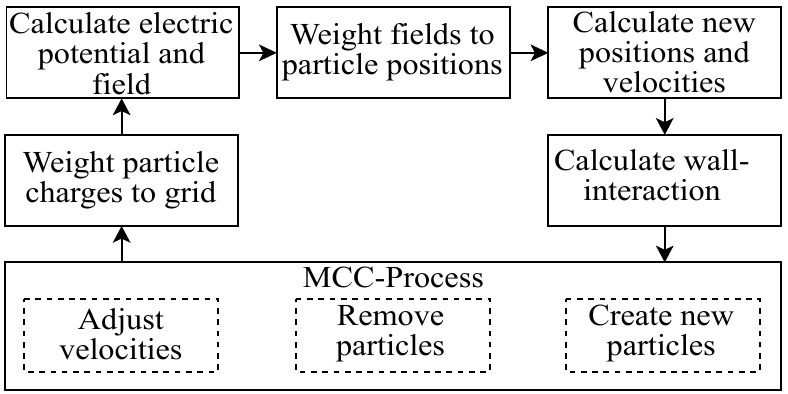}
\caption{Schematic illustration of the simulation cycle during one time step $\Delta t$ iterated until simulation convergence.}
\label{fig:schematic_pic}
\end{figure}

The particle-in-cell (PIC) method conceptually relies on the previously introduced Monte Carlo method.\cite{birdsall_plasma_1985,hockney_computer_1988,turner_simulation_2013} As suggested by its name, a division of the computational (spatial) domain into discrete grid cells is exploited. Specifically, in the one dimensional setup depicted in figure~\ref{fig:setup}, the discharge gap $L$ is discretized into $N$ segments of length $\Delta x$.

The evolution of the system is traced in an iterative evaluation cycle.\cite{birdsall_particle--cell_1991} A schematic of one cycle of duration $\Delta t$ is given in figure~\ref{fig:schematic_pic}. In brief, the cycle starts with an assignment of the individual super-particle charges and currents to the nodes of the spatial grid. From the charge and current density at the grid points the electric (and magnetic field) is computed. In order to push the particles according to the acting forces, actually the Lorentz force, the fields first have to be interpolated back to the individual particle positions. They are subsequently pushed in configuration and velocity space according to Newton's laws. As particles traverse through the spatial domain and reach the walls, their interactions with these walls have to be accounted for. Finally, collisions are taken into account by means of Monte Carlo collisions (MCC) and the cycle starts over. Although previously described quite generally, in the following considerations the problem complexity is reduced by the absence of both externally applied and self-consistent magnetic fields. Consequently, the electrostatic approximation of Maxwell's equations is justified, leading to Poisson's equation, which has to be solved within each cycle.

\paragraph{Charge Assignment.}
The conceptual beauty of the PIC method lies within the treatment and decoupling of charge interaction processes. Charged particles do not only experience short range interactions (i.e., elastic or inelastic collisions), but also long range Coulomb forces. In principle, these can be incorporated into the simulation by direct evaluation of the individual interactions for each pair of charged particles respectively, ignoring self-interactions. Computationally, however, this leads to the dilemma of unfeasible computational requirements. Within PIC, rather than to directly evaluate the individual Coulomb interactions, by appropriate spatial mapping and interpolation self-interaction is circumvented relaxing the coupled solution problem to a feasible task -- following the Clouds-in-Cells idea.\cite{birdsall_clouds--clouds_1969} To obtain the charge density $\rho$ at the grid nodes, the charges of the ions and electrons given at the individual particle positions $\vec{r}_k$ need to be weighted onto the discrete grid nodes, for instance the $m$-th node at $x_m = m \Delta x$. Different weighting methods are possible.\cite{birdsall_clouds--clouds_1969,birdsall_plasma_1985} One of the simplest energy bounded interpolation schemes is a linear weighting.\cite{birdsall_clouds--clouds_1969}. In one dimension, first a triangle of width $2\Delta x$ and centered at the continuous particle position is considered. The super-particles point charge is distributed onto the grid nodes covered by the triangle -- due to its limited width at most two nodes. By ensuring that the total charge distributed equals the charge of the super-particle $q_k$, the grid based charge density is obtained.

\paragraph{EM Field Solver and External Lumped Circuits.}
For the calculation of the forces acting on the super-particles, the electric field, and if necessary also the magnetic field, has to be determined. In addition to the internal forces, technological plasmas are almost exclusively driven via externally supplied power. The corresponding forces need to be included in the simulation efficiently, as outlined in this paragraph at the example of the electric field.\cite{birdsall_plasma_1985,verboncoeur_simultaneous_1993} The electric potential and electric field are present on the same grid as the charge density. To calculate the fields, Poisson's equation $\nabla^2\Phi (\vec{r})=-\rho (\vec{r})/\epsilon_0$ is discretized with a second order finite difference scheme and solved numerically given the charge density $\rho (\vec{r})$. In the one dimensional case, this is $(\Phi_{m+1}-2 \Phi_m+ \Phi_{m-1})/\Delta x^2=-\rho_m/\epsilon_0$, where $m$ indicates the respective grid point. From $\Phi (\vec{r})$, or $\Phi_m$ respectively, the electric field is easily obtained via $\vec{E}(\vec{r})=-\nabla \Phi (\vec{r})$, in the one dimensional case via $E_m = -(\Phi_{m+1}-\Phi_{m-1})/2\Delta x$. The external excitation applied to the electrodes specifies the boundary conditions through their surface charges $\sigma$. This holds if the source is connected via an external lumped circuit. The surface charge $\sigma$ is used as the interface connecting the plasma and the external circuit. Simply speaking, the boundary condition is chosen in a way that the continuity equation is implicitly satisfied throughout the simulation. For the specific details of the implementation, reference \cite{verboncoeur_simultaneous_1993} is strongly recommended.

\paragraph{Field Interpolation.}
Once the electric (and if necessary magnetic) field has been obtained, the forces acting on the ensemble of super-particles have to be taken into account. As the fields and corresponding forces are known on the numerical grid only, an intermediate interpolation is required to obtain the forces at the specific super-particle positions $\vec{r}_k$. Again, a straightforward approach is the linear interpolation as previously mentioned in the context of charge assignment.\cite{birdsall_clouds--clouds_1969} The formalism in one dimension is principally the same: taking a triangle of width $2 \Delta x$ centered at the position of the super-particle, the force from the two adjacent nodes is averaged (inversely weighted by distance to the grid points $\Delta x$).

\paragraph{Particle Pushing.}
The method of how to apply the forces on the particles and to calculate their new positions and velocities depends on specific situation. If a magnetic field is involved, the particle pusher developed by Boris has proved to be efficient.\cite{boris_relativistic_1970} In the absence of a magnetic field the situation simplifies greatly. The new position can be calculated using \eqref{eq:pushinr}. The velocity $\vec{v}_k^{\, (l+1/2)}$ is updated based on \eqref{eq:pushinv} with acceleration $\vec{a}_k^{\,(l)} = (q_k/m_k) \vec{E}(x_k^{\,(l)})$, directly obtained from the Lorentz force. The new position $\vec{r}_k^{\, (l+1)}$ is then calculated. In a one dimensional case, position vector $\vec{r}_k^{\, (l+1)}$ is replaced by the scalar position $x_k^{\, (l+1)}$.

\paragraph{Wall Interaction.}
Depending on the case of interest the walls may act as particle sinks, reflect particles or -- as in the exemplary setup -- constitute a source for sputtered particles. If the latter is the case, the energy and angle of the incoming ions are essential, once they reach the wall. These values are taken as input for the computation of the sputtering yield and distribution. The details of which are covered in subsection~\ref{sec:tpm}. The ion itself is almost exclusively neutralized in the process of a wall interaction and contributes to the neutral background. That is, given a constant background density, it is simply removed from the simulation. The same is done with wall interactions for which the walls act as perfect particle sinks, as for aluminum particles in the exemplary setup (whether ions or not). When particle reflection has to be taken into account and the (energy dependent) reflection probability $P_\mathrm{refl}$ is known, a simple approach is to compare $P_\mathrm{refl}$ with a uniform random number $R_\mathrm{f}$. If $R_\mathrm{f} \geq P_\mathrm{refl}$, the particle is not reflected and removed from the simulation. Otherwise, the reflection is calculated. For the easiest case of a specular reflection (typically a good approximation for clean metal surfaces), the velocity component perpendicular to the wall is simply inverted. Another effect which may be important is secondary electrons emission, which principally can be taken into account similarly on a probabilistic basis. Various particles (e.g., ions, electrons, excited species, or energetic neutrals) may give rise to this phenomenon. However, for the sake of simplicity, it is neglected in the present case. 

\paragraph{Particle Collisions.}
An aspect of major importance for kinetic simulation methods like PIC are the collisions of particles and the way they are incorporated in the simulation framework.\cite{birdsall_particle--cell_1991} A fundamental factor defining the collision dynamics is the energy dependent cross section $\sigma(\epsilon)$, where the interaction energy $\epsilon$ is related to the relative speed as $\epsilon = 0.5 \, m_\textrm{r} \, v_\textrm{r}^2$, with $m_\textrm{r} = m_1 m_2/(m_1 + m_2)$ being the reduced mass. Within PIC models the cross section is usually utilized in the form of a look-up table from experimental measurements. Depending on the kind of collision process of interest (e.g., ionization, charge exchange, elastic scattering) and the involved species, different data sources for the cross section are available in the literature, particularly in the LXcat database.\cite{_http://www.lxcat.laplace.univ-tlse.fr_2016}

For weakly ionized low temperature plasmas, the neutral gas in the vacuum chamber is most abundant. Hence, only collisions of the various plasma species with the background gas are to be taken into account. It is computationally exceptionally expensive to repetitively calculate the collision probability for each possible kind of process individually. More cost effective approaches have been proposed -- specifically, the null collision method.\cite{skullerud_stochastic_1968,lin_null-event_1978} From the definition of a cumulative energy dependent collision cross section $\sigma_\mathrm{sum}(\epsilon) = \sum\nolimits_{j=1}^J \sigma_j (\epsilon)$, summed over all energies a maximum collision frequency
\begin{align}
	\nu_\mathrm{c,max} = \left[ n_\mathrm{bg} \, \sigma_\textrm{sum}(\epsilon) \, \sqrt{2 \, \epsilon / m_\textrm{r}} \right]_\textrm{max}
\end{align}
is obtained. Subsequently, using equation \eqref{eq:collisionprobability1} the energy independent total collision probability $P_\mathrm{c,tot}$ is calculated (once during initialization of the simulation). This corresponds to an upper probability bound. Throughout the simulation, irrespective of the actual interaction energy $\epsilon$, collisions are taken into account and evaluated with probability $P_\mathrm{c,tot}$ given a fixed $\Delta t$ for each time step as described previously. Whenever a collision takes place, the type of collision has to be selected as well. For the specific interaction energy of concern, the cumulative cross section $\sigma_\mathrm{sum}(\epsilon)$ is virtually complemented with a so-called null-collision process $\sigma_\textrm{null}(\epsilon)$ so that $\nu_\mathrm{c,max}\equiv\nu_\textrm{c,sum}(\epsilon) + \nu_\textrm{null}(\epsilon)$. A random collision sample is obtained by multiplication with an new random sample $R_\textrm{f}$, so that $0 \leq \nu_\mathrm{rand} \leq \nu_\mathrm{c,max}$. Since every possible collision is included, the value of $\nu_\mathrm{rand}$ selects the $l$th type collision process -- essentially from $\sum\nolimits_{j=1}^{l-1}\nu_j(\epsilon) \leq \nu_\mathrm{rand} \leq \sum\nolimits_{j=1}^{l}\nu_j(\epsilon)$. Of course, for collisions involving an energy threshold (e.g., ionization or excitation) the interaction energy has to be checked additionally. In case a null collision event is chosen, the evaluation is stopped and no collision takes place.

\paragraph{Limitations and Constraints.}
It is important to have in mind the conceptual assumptions and limitations. For the explicit evaluation scheme described within this manuscript, the size of one grid cell $\Delta V$ is subject to limitations. In order to be energy bound, the numerical cell dimensions are limited to $\Delta V^{1/3} \ll \lambda_{\mathrm{D}}$ (i.e., much smaller than the Debye length).\cite{birdsall_plasma_1985} To ensure stability, the choice of the time step is constrained to $\Delta t \ll \omega_{\mathrm{p}}^{-1}$, with the electron plasma frequency $\omega_{\mathrm{p}}=(n_\mathrm{e}e^{2}/\varepsilon_0 m_\mathrm{e})^{1/2}$. Electrons are typically much faster than ions, which is why they are the limiting factor regarding the time step. Different evaluation schemes have been proposed, relaxing the mentioned constraints, in particular relying on an implicit evaluation of the PIC cycle.\cite{cohen_implicit_1982,langdon_direct_1983,friedman_second-order_1990,vahedi_capacitive_1993,chen_analytical_2013}

The number of simulated super-particles is a different matter. Depending on the dimensionality of the simulation and the physical environment, the particle weight has to be defined. Common definitions include the number of simulated particles per Debye length $N_{\mathrm{D}}$ and per grid cell $N_{\mathrm{C}}$. Since a proper grid cell size $\Delta V$ also depends on $\lambda_{\mathrm{D}}$, $N_{\mathrm{C}}$ is a good parameter of choice. A good estimation for many applications is to set the particle weight so that $N_{\mathrm{C}} \geq 20$.\cite{turner_simulation_2013}
 
Different implications come into play when considering cases with reduced dimensionality (e.g., one dimensional as in this tutorial). Firstly, within the Poisson solver, three dimensional point charges translate to planes of charges in a one dimensional evaluation. Invariance in lateral direction is inherent to the simulation, implying a different physical behavior. Moreover, even in situations where a one dimensional representation in configuration space seems applicable, in velocity space it may not. The most prominent choice, therefore, is to reduce dimensionality in configuration space, but fully maintain a three dimensional velocity space. Thereby, particle positions $\vec{r}_k$ translate to a scalar position $x_k$, while the three dimensional velocity $\vec{v}_k$ is preserved. In particular, throughout collisional processes scattering into different directions in velocity space is accounted for. The model is consequently referred to as ``1D3V''.

\subsection{\label{sec:tpm} Test Particle Method}
Neutral particles within the plasma may be treated separately from the charged particle contribution. They are, however, coupled via the input from the ion flux and energy distribution at the surfaces. In the given scenario, a kinetic transport description is required due to the comparably large Knudsen number. One should keep in mind that the sputtered particles have an inherently non-thermal velocity distribution. A kinetic description is therefore advisable. The neutral particle model is concerned with solely the collisional transport of energetic sputtered particles. Self-interaction (i.e., metal with metal) is neglected, as arguable by the condition of a constant background gas and only a trace of aluminum therein. The individual aluminum interactions with the argon background are assumed to be of binary nature. Due to the low pressure three-body collisions can be safely neglected. For this situation, the test particle method (TPM) is most appropriate. The heavy particle dynamics can be efficiently ``decoupled'' from the fast electron and plasma dynamics. Principally, TPM can be also applied to the general case of ion transport taking into account the interaction of the particles with the background atoms as well as with a known electric field (e.g., provided from PIC simulations).\cite{minea_kinetics_2014} This shall not be detailed in this work. As a major prerequisite, it is important to notice that any type of feedback onto the plasma is inherently omitted then.

Regarding the methodology various variants are documented in the literature, which have been widely applied over decades for the simulation of transport phenomena and gas kinetics. \cite{somekh_thermalization_1984,motohiro_monte_1984,turner_monte_1989,myers_monte_1992,turner_monte_1992,clenet_experimental_1999,van_aeken_metal_2008,depla_magnetron_2012,lundin_tiar_2013} A peculiar variant is the test multi particle method (TMPM).\cite{kersch_selfconsistent_1994,serikov_monte_1996,trieschmann_transport_2015} The general beauty of TMPM again lies within its conceptual separation from the physical processes under consideration. That is, book keeping and numerical handling is one rather technical aspect, the physical considerations are a different one.

\paragraph{Collisional Transport.}
The previous considerations can be applied to the case of energetic neutral particle transport subject to binary collisions within a vacuum environment. For the generalized situation of an inhomogeneous force field, a constant time step $\Delta t$ is suggested. Based on the previous considerations and in particular the ensemble averaged expectation value for the collision frequency the transport can be described. On the one hand, taking equation \eqref{eq:collisionprobability1} and linearizing with $\tau \equiv \Delta t$ and $\tau_\textrm{c} \equiv 1/\nu_\textrm{c,mean}$ yields
\begin{gather}
	P_{c} = n_\textrm{bg} \, \overline{\sigma(v_\textrm{r}) v_\textrm{r}} \, \Delta t. \label{eq:collisionprobability2}
\end{gather}
On the other hand, given a Lagrangian collection of super-particles, the ensemble average is equivalently evaluated by exercising the calculation of the collision probabilities for all particle individually. In principle, the most imperative constraint is that $\Delta t$ be small enough, so $P_{k,\textrm{bg}} \ll 1$. This concept in general works quite well.\cite{kersch_selfconsistent_1994,serikov_monte_1996} Problematically, the above mentioned constraint renders the evaluation quite inefficient. For improved efficiency the no-time counter scheme can be an alternative. This shall not be the focus of this manuscript, but is detailed elsewhere.\cite{bird_perception_1989,bird_molecular_1994,kersch_transport_2011,trieschmann_transport_2015}

\paragraph{Binary Interactions.}
For evaluation of the individual binary interactions, the differential collision cross section $\sigma_\textrm{d}(v_\textrm{r}, \chi)$ is important. Its velocity and angular dependence differs depending on the nature of the collision. For practical purposes, oftentimes the total collision cross section $\sigma(v_\textrm{r}) \equiv \sigma_\textrm{T}(v_\textrm{r}) = 2\pi \int_{0}^{\infty} b(v_\textrm{r}) \, d b$ is evaluated. Two premises are: Firstly, a finite distance of interaction so that the upper bound of the integral is $b_\textrm{max}$ and $\sigma_\textrm{T}(v_\textrm{r}) = \pi b_\textrm{max}^2(v_\textrm{r})$ irrespective of scattering details. And secondly, the details of the interaction depend on a specific scattering law (i.e., even a large total cross section may lead to only small angles of deflection). In order to be applicable in theoretical models, the cross section has to be numerically known (may it be in the form of an analytic formula or numerical look-up tables). Analytic expressions have the advantage of well known scattering dependencies. A drawback is that barely all the details of a scattering interaction are captured. The collision process is best viewed in the co-moving frame of the collision.\cite{vincenti_introduction_1967,bird_molecular_1994}

The most prominent choice is the variable hard sphere (VHS) model\cite{bird_monte-carlo_1981} with an isotropic distribution of the scattering angle $\chi(b) = 2\arccos\left( b / b_\textrm{max} \right)$ on a unit sphere -- a function of the impact parameter.\cite{bird_molecular_1994} Due to its isotropic angular distribution, the VHS model is easily transferred into the laboratory frame (i.e., it is isotropic in either coordinate system) thus requiring no coordinate transformation. An energy (and thus velocity) dependence is commonly implied following a power law for the corresponding viscosity $\mu(T) = \mu(T_\textrm{ref}) (T/T_\textrm{ref})^{\tilde{\omega}}$, specified for combinations of reference temperature $T_\textrm{ref}$ and viscosity index $\tilde{\omega}$ (e.g., fitted to experimental data \cite{bird_molecular_1994,morokoff_comparison_1998,kersch_transport_2011,trieschmann_transport_2015}). From these considerations, the maximum impact parameter can be defined\cite{bird_molecular_1994,kersch_transport_2011}
\begin{align}
	b_\textrm{max}(v_\textrm{r}) &= d_\textrm{ref} \, \sqrt{ \frac{1}{\Gamma(2.5-\tilde{\omega})} \, \left( \frac{2 k_\textrm{B} T_\textrm{ref}}{m_\textrm{r} v_\textrm{r}^2} \right)^{\tilde{\omega}-0.5}}. \label{eq:maximumimpactparameter}
\end{align}
In an alike collision event of two particles of species $\alpha$ this can be directly evaluated using $m_\textrm{r} = m_\alpha / 2$ and the reference diameter for this species
\begin{align}
	d_\textrm{ref} &= \sqrt{\frac{15 \, \sqrt{ m_\alpha k_\textrm{B} T_\textrm{ref} / \pi}}{8 (2.5-\tilde{\omega})(3.5-\tilde{\omega}) \, \mu(T_\textrm{ref})}}. \label{eq:referencediameter}
\end{align}
For an unlike collision expression \eqref{eq:maximumimpactparameter} can be employed using and a linear interpolation $d_\textrm{ref} = (d_{\textrm{ref},\alpha}+d_{\textrm{ref},\beta})/2$ and $\tilde{\omega} = (\tilde{\omega}_{\alpha}+\tilde{\omega}_{\beta})/2$ evaluating (\ref{eq:referencediameter}) for species $\alpha$ and $\beta$, respectively.

More complicated choices for the collision model are available, for instance the M1 collision mode, a modified VHS model.\cite{kersch_selfconsistent_1994,morokoff_comparison_1998,serikov_monte_1996}). These models typically share an anisotropic scattering nature. For the M1 model this is $\chi(b) = \pi \left( 1 - b/b_\textrm{max} \right)$, anisotropically distributed over a unit sphere. Regarding the energy scaling, equations \eqref{eq:maximumimpactparameter} and \eqref{eq:referencediameter} can be readily employed. The reference data available in the literature for VHS can be used for M1 with the reference diameter $d_\textrm{ref,M1} = \sqrt{4/3} \, d_\textrm{ref,VHS}$.\cite{morokoff_comparison_1998} Not of concern in this work, but relevant when analytic models are favored instead of look-up tables, are theoretical approaches to collisional interactions of charged particle. An illustrative approach is the one originally published by Langevin in which the charge induced interaction is consistently taken into account.\cite{langevin_formule_1905,hasse_langevins_1926} A detailed study of which and its implementation has been proposed by Nanbu et al.\cite{nanbu_ion-neutral_1995,nanbu_probability_2000}

In practical terms either collision model requires for the evaluation of a random collision event a choice of the impact parameter. This is to be sampled from a random distribution. A probability density uniformly distributed over a cross section with the substitution $B=b^2$ and $dB=2b \, db$ implies
\begin{align}
	p_\sigma d\sigma = \frac{dB d\psi}{\sigma_\textrm{T}} =  \frac{2b\, db d\psi}{\sigma_\textrm{T}}.
\end{align}
The probability density in terms of the impact parameter $b$ and the azimuthal angle $\psi$ is therefore linear in $b$ (consistent with a Boltzmann collision kernel). The corresponding impact parameter distribution is sampled by the method of inversion from $b_0 = b_\textrm{max} \, \sqrt{R_\textrm{f}}$.\cite{bird_molecular_1994} To fully specify the scattering direction also a random polar angle $\psi_0 = 2 \pi R_\textrm{f}$ is chosen. In order calculate the post-collision velocity of the scattered particles in the laboratory frame in the anisotropic situation, also a consecutive coordinate transform is required.\cite{bird_molecular_1994,kersch_transport_2011}

\paragraph{Wall Interactions.}
An aspect not detailed in subsection~\ref{sec:pic} concerns the implications of the plasma-wall interaction for the neutral particle model -- more precisely, its manifestation in terms of a sputtered particle flux. It is the imperative component regarding the coupling from the PIC to the TPM model. The sputtered particle flux (i.e., the number of particles emanated from the surface per unit area and unit time) is most easily realized for super-particles of same weight within PIC as well as TPM. In detail, rather than to a priori specifying the total number flux of sputtered particles, for each ion hitting the wall the interaction can be evaluated on a single event basis. That is, for every sputtering event the probability can be calculated from a given sputtering yield $Y(E_0,\theta_0)$ -- a function of the projectile's energy $E_0$ and its angle of incidence $\theta_0$. For a vast amount of projectile-target combinations fitted parameters are available from Behrisch and Eckstein. These parameters can be used to evaluate the respective analytic expressions.\cite{behrisch_sputtering_2007,eckstein_sputtering_2008} The sputtering yield can be interpreted from a probabilistic perspective: If an ion fulfills the initial criteria to perform a sputtering event, a particle is sputtered given that $R_\textrm{f} \leq Y$. On the occasion that the sputter yield is larger than one, the check can be performed $N$ times with $R_\textrm{f} \leq Y/N$ and $N = \textrm{ceil} \left( Y \right)$. For each successful check, a new particle is inserted into the volume with an initial position equal to the location of ion impact.

In order to fully specify a sputtered particle to be inserted, its velocity and angle of emission are to be known. These may be conveniently sampled from a given particle distribution. Notably, however, what is actually specified is the velocity distribution $f(\vec{v})$ of the particle flux, not the velocity distribution of the particle density (within a given volume). Both of which are indeed related by a scaling with the normal component of the velocity $\vec{n}\cdot\vec{v}$. In the case of sputtering it is important to acknowledge that a sputtered particle behaves vastly different than what is to be expected from a thermal influx. The latter being significantly more energetic with energies on the order of eV, in contrast to thermal particles with typical energies of $25$~meV. As such, sputtered particles are commonly assumed to follow a Sigmund-Thompson distribution.\cite{thompson_ii._1968,sigmund_theory_1969}. Several variants are published in the literature.\cite{serikov_monte_1996,stepanova_estimates_2001,trieschmann_transport_2015} The details depend on the specific situation. For a more accurate description, the output from transport of ions in matter (TRIM) or even better molecular dynamics (MD) simulations may be readily employed.\cite{eckstein_computer_1991,urbassek_molecular-dynamics_1997,eckstein_sputtering_2008}

The individual particle properties may be obtained from an acceptance-rejection sampling of a given energy distribution, paired with an emission angle obtained from sampling of a correspondingly given angular distribution.\cite{bird_molecular_1994} The most common choice is the assumption of a cosine distribution $f(\theta) \sim \cos \left( \theta \right)$ and therefore $\theta=\arccos\left( \sqrt{R_\textrm{f}} \right)$.\cite{knudsen_cosinusgesetz_1916,turner_monte_1989,greenwood_correct_2002}

When a neutral particle impinges a wall, several pathways are possible. It may be specularly reflected, be thermally re-emitted, or stick to the wall. While the former two choices are common for inert gas flows, the latter is most relevant for the scenario analyzed in the paper.\cite{hansen_atomistic_1999} In this case, the walls can be envisioned as perfect particle sinks and any particle approaching is removed from the simulation. The absorbed flux, however, is highly relevant for the purpose of a later analysis.

\paragraph{Limitations and Constraints.}
Throughout a TPM simulation, a few constraints have to be accounted for. The foremost important parameter concerning collisional transport is the discrete time step $\Delta t$. It is required to be chosen small enough to properly account for all collisions encountered. That is, given at hand equation \eqref{eq:collisionprobability2} governed by the expectation collision time $\tau_\textrm{c}$, a correct physical representation is ensured only, when $\Delta t \ll \tau_\textrm{c}$ is satisfied.

In the inhomogeneous situation $\Delta t$, moreover, has to be chosen small enough to properly resolve any local gradients. This is straightforwardly linked to the distance a particle travels during one iteration. When a mesh with cell size $\Delta V$ is employed, at all times this distance shall not exceed $\Delta V^{1/3}$ and therewith any given gradient length scale.\cite{sun_proper_2011}

In terms of the diagnostics, it is important to distinguish the various TPM simulation types. In grid based TPM simulations, the number densities may be forthrightly obtained from a cell based average of the particles residing within. In contrast, within gridless simulations with adaptive time steps $\Delta t \equiv \tau_\textrm{c}$, the particle density is obtained taking into account the specific ``flight duration''. A time weighted evaluation is obligatory.\cite{turner_monte_1989}

\section{\label{sec:results} Simulation Results}

\begin{figure}[t!]
\includegraphics[width=8cm]{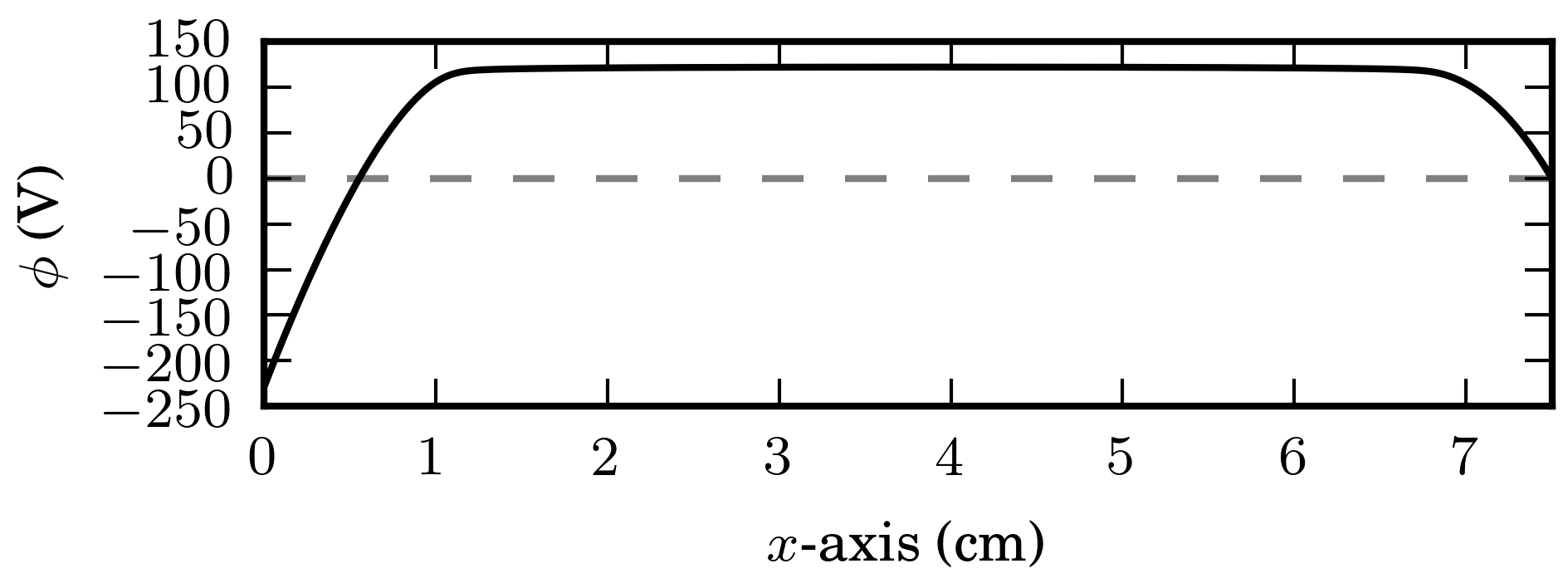}
\caption{Spatial profile of the average electric potential obtained from the coupled PIC-TPM simulation for the specified parameters.}
\label{fig:results_potential}
\end{figure}

\begin{figure}[t!]
\includegraphics[width=8cm]{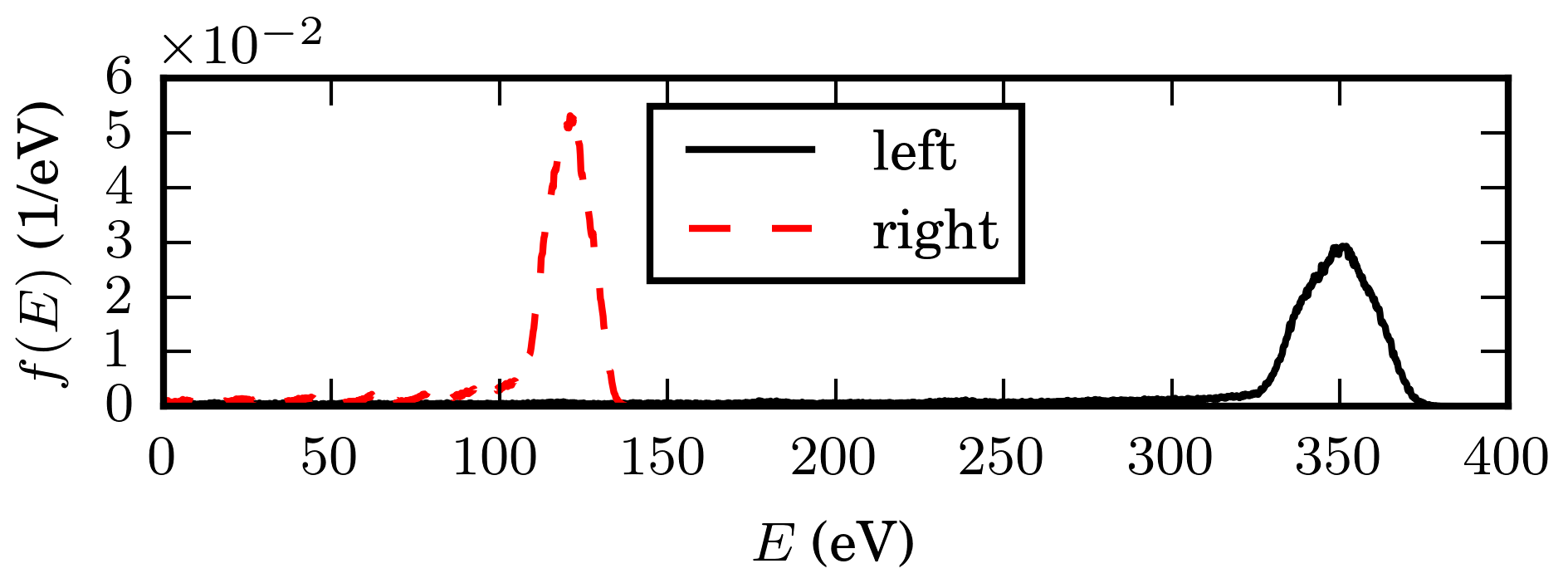}
\caption{Ion energy distribution functions of ions impinging the target (left; solid black) and substrate (right; dashed red) electrode, respectively.}
\label{fig:results_iedf}
\end{figure}

\begin{figure}[t!]
\includegraphics[width=8cm]{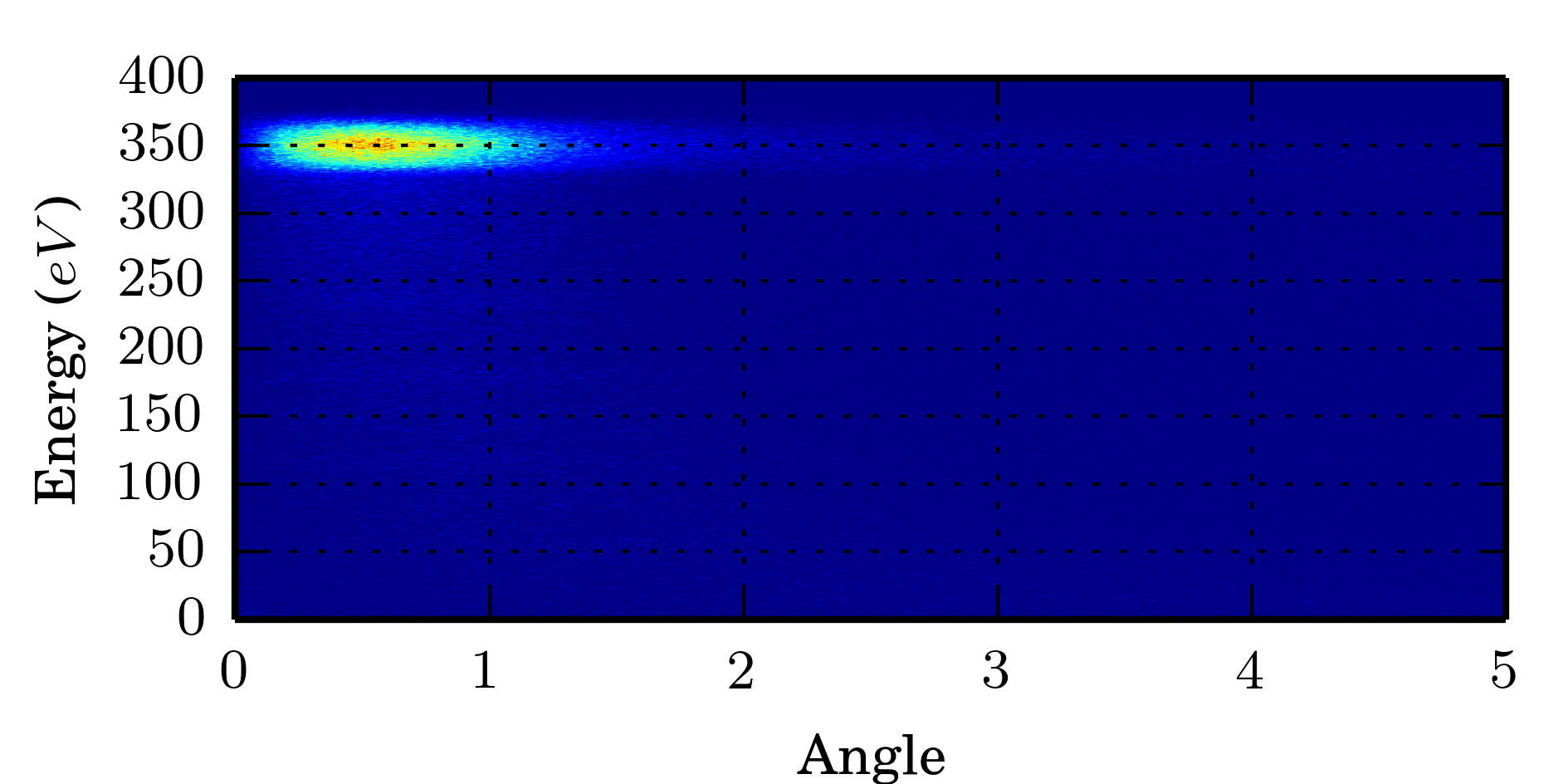}
\caption{Contour plot presenting the energy and angular distribution of ions impinging the target. Colorscale: Blue represents zero, red maximum.}
\label{fig:results_ieadf}
\end{figure}

\begin{figure}[t!]
\includegraphics[width=8cm]{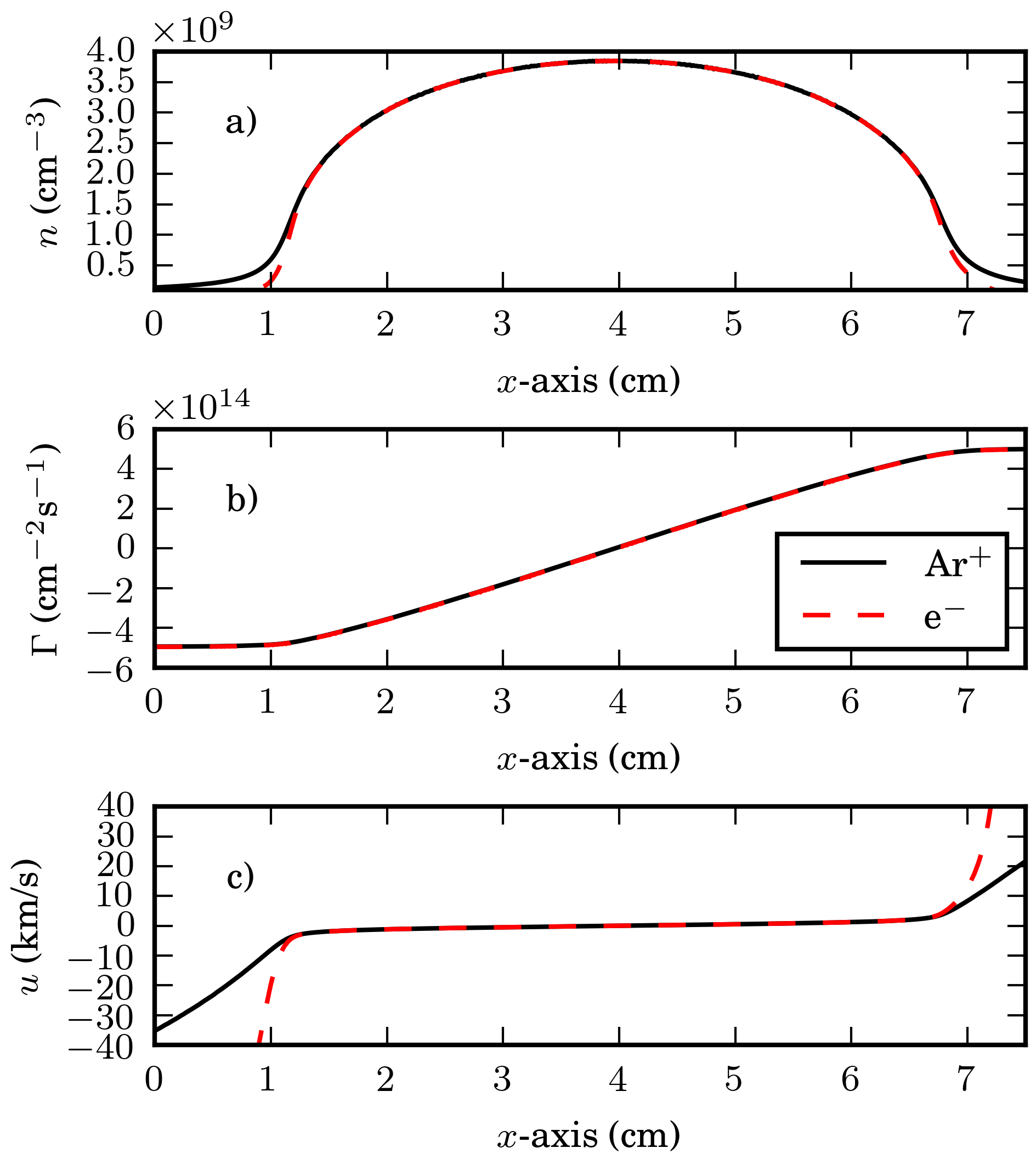}
\caption{a) Temporally averaged spatial distribution of the electron (dashed red) and argon ion (solid black) density. b) Corresponding average electron and ion flux. c) Respective average mean velocities calculated for both electrons and ions.}
\label{fig:results_densities}
\end{figure}

\begin{figure}[t!]
\includegraphics[width=8cm]{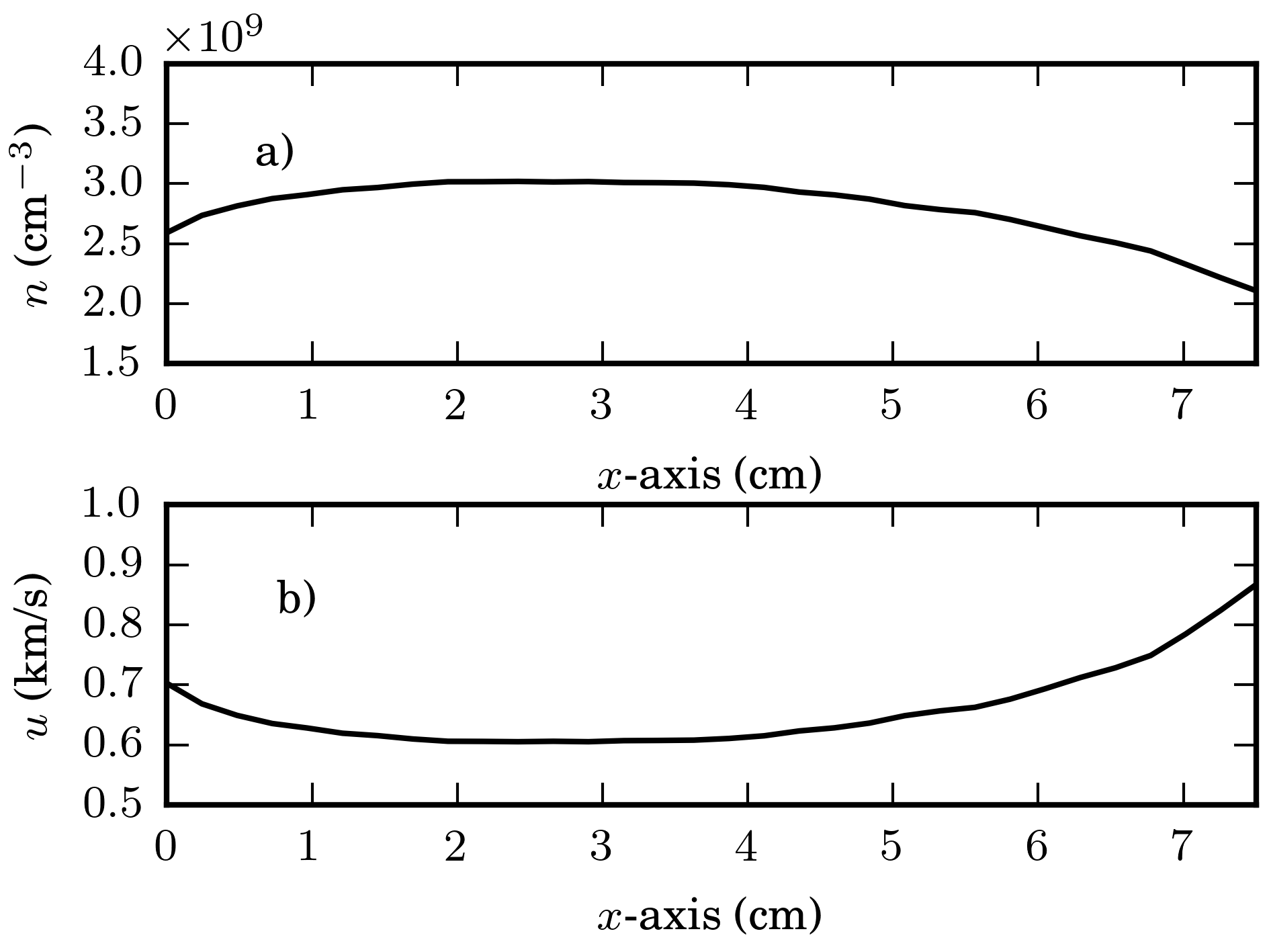}
\caption{Spatial profiles of sputtered aluminum within the discharge gap: (a) Average density and (b) the corresponding average mean velocity.}
\label{fig:results_metal}
\end{figure}

Equipped with the details specified in the preceding sections a brief review of some exemplary results is an instructive task. Figures~\ref{fig:results_potential} to \ref{fig:results_metal} show simulation results for the exemplary case specified in section~\ref{sec:setup}. PIC is used for the plasma simulation, whereas a consistently coupled TPM is used for the sputtered neutral species. The average electric potential is plotted in Figure~\ref{fig:results_potential}. As is to be seen therein, a DC self-bias voltage of $\approx 225$~V establishes due to the electrically asymmetric excitation of the plasma. More importantly, in front of both electrodes a substantial voltage drop is found, approximately $350$~V (left), $125$~V (right). Consequently, ions which undergo acceleration in this sheath potential more or less permanently impinge on the electrodes with substantial energies and may possibly sputter the electrode surface. This is important for a twofold reason: Firstly, the main achievement is the sputtering of the target surface (left). Secondly, also re-sputtering of the substrate surface (right; initially stainless steel, but in fact aluminum coated) is to be taken into account. For both electrodes, the corresponding ion energy distributions (IEDs) of argon ions impacting on the electrodes are shown in figure~\ref{fig:results_iedf}. The IEDs are clearly centered around mean energies corresponding to the respective voltage drop in front of the individual electrodes. To a certain degree, the IEDs are spread around the peak value, which can be argued from the fact that also ions do not truly experience the average electric field and potential, but to some extent the time varying. (Electrons respond to the transient electric field in any case.) As such, ions are slightly modulated as well, as there are times when the potential drop in front of the electrodes is smaller or greater than the averaged value.\cite{benoit-cattin_anomalies_1968,metze_energy_1989,mussenbrock_modeling_2012} In addition, a low energy plateau with periodic humps is vaguely observed for both energy distributions. These are known to be due to elastic resonant charge exchange collisions.\cite{wild_ion_1991}


The quantitative aspects of the plasma dynamics are largely determined by processes taking place in the gas phase. Of particular interest are the mean plasma quantities, that is, the mean electron and ion densities, their mean fluxes, and velocities. Each of which is depicted in figure~\ref{fig:results_densities}. Several points are important: (i) technological radio-frequency plasmas are predominantly driven by selective external energy coupling into the electron component. Correspondingly, their density and energy distribution is of great interest. Quasineutrality typically establishes within the plasma volume and so the ion and electron density roughly equalize in the bulk of the plasma. As a function of distance, the average densities are presented in figure~\ref{fig:results_densities}a). A characteristic charge depletion in the boundary sheath can be clearly identified. (ii) The particle fluxes to the walls -- one of the most influential parameters governing the sputtering process -- follow accordingly. In steady state operation, on average the volumetric generation and loss terms balance. That is, under the assumption of a Bohm flux leaving the plasma bulk and entering the bounding plasma sheath, the fluxes of electrons and ions are on average equal. This is also observed from figure~\ref{fig:results_densities}b). It is important to keep in mind that both, electrons and ions, in principle experience a time varying electric field. Yet, while electrons are governed by the fast transients (i.e., they react almost instantaneously), ions are mostly affected by the time average. In terms of quantitative considerations, the ion flux (i.e., the number of ions per unit area and unit time) impinging the target has strong implications for the flux of sputtered metal originating from the very given surface. (iii) In this respect another vastly influential parameter is the mean velocity of the incident ions, displayed in figure~\ref{fig:results_densities}c). It is inherently linked to the IED formerly addressed. A decisive difference is that the mean velocity is an example of a volumetric quantity, while the IED corresponds is a fluxual property. Of relevance for the sputtering process, however, is solely the latter in particular at the plasma-wall interface.

\paragraph{Sputter Deposition}
When the peculiar focus is the understanding of the sputter deposition, the described model approach requires a distinctive coupling of the plasma related properties into the neutral particle model. The flux of sputtered particles originating from the surfaces manifests in a steady state spatial profile of the aluminum density and mean velocity consistently obtained within the coupled PIC-TPM simulation. This is depicted in figure~\ref{fig:results_metal}. The flux of particles is not of primary interest as particles are assumed to be solely generated at or lost to the walls. As such the metal flux is continuous throughout the discharge volume. (This is different for the plasma, as in this case electron-ion pairs are generated in the bulk and lost to the walls. Nearly no ionization takes place in the boundary sheath, which manifests in a straight line in figure~\ref{fig:results_densities} indicating flux continuity.) The distinct profiles of the sputtered species' density and velocity stem from the interaction of particle fluxes originating from the target (left to right) as well as re-sputtered from the substrate (right to left). Both electrodes act as particle sources and sinks. As such, minima in the density are observed near the boundaries. Of particular importance in this respect is the collisional transport through the discharge gap. Particles suffering a collision are isotropized (i.e., may be forward or backward scattered). As the sputtered particle flux directed left to right is larger than right to left, the mean density and velocity profiles become asymmetric. Due to the backscattered component and the requirement for flux continuity, the velocities are observed to increase adjacent to the surfaces. This can be understood from the details of the particle distributions within the discharge gap. Details can be found for example in references \cite{turner_monte_1989,trieschmann_transport_2015} and shall not be reconsidered at this point.

\paragraph{Hypotheses validation}
Finally due is a discussion of the previously in section~\ref{sec:setup} introduced assumptions often used in the context of sputter transport simulations: Assumption (A) can be readily verified from the ion energy distribution. As displayed in figure~\ref{fig:results_iedf}, ions do not hit the surfaces strictly mono-energetically. However, considering the width-to-peak ratio $\Delta E / E_\text{max} \approx 0.1$ at both the target and the substrate, the spread in energy is negligible. It is due to collisions as well as the radio-frequency modulation. In conclusion, mono-energetic impingement is a good approximation. Regarding the angular distribution figure~\ref{fig:results_ieadf} suggests an angular dispersion of less than 1 degree off surface normal, as easily verified. 

Assumption (B) can only indirectly be validated. That is, the proposed model does not permit to justify whether the influence of the plasma and the energetic neutral particles is irrelevant. In contrast, what can be verified is the contents of the respective species as quantified by their partial densities in comparison with the argon background gas. On the one hand, the degree of ionization is $n_\text{Ar+}/n_\text{Ar} \lesssim 10^{-4}$ and rendering the corresponding rarefaction insignificant. On the other hand, the neutral aluminum to neutral argon fraction is even lower with $n_\text{Al}/n_\text{Ar} \lesssim 5 \times 10^{-5}$ and gas rarefaction due to sputtering wind effects is immaterial as well. Taking into account ionization of the metal as well, the density of singly-ionized aluminum $n_\text{Al+}$ (not depicted, but incorporated into the simulation) is roughly a factor 1800 smaller than the argon ion density $n_\text{Ar+}$ and thus absolutely obsolete.\cite{shimon_effective_1975}

In conclusion, both assumptions previously postulated and exploited throughout this paper are validated by corresponding observations in the obtained simulation results. The reduced model (figure~\ref{fig:interactions}) is unconditionally justified.

\section{\label{sec:conclusion} Conclusions}
The aim of this manuscript is to provide an audit of simulation approaches to low pressure plasma discharges coupled to a dynamic neutral particle component. In particular, this manuscript presents a means of guidance to the reader: (i) by providing basic access to the topic for some concerned with the implementation of such model, or (ii) by opening up the perspective for others in order to have a feeling about the complexity and limitations of such model approaches. Starting from a critical statement of the scenario provided in section~\ref{sec:setup}, a brief introduction into Lagrangian based Monte Carlo modeling is provided. In section~\ref{sec:mcsimulations}, at the example of a radio-frequency driven capacitive sputtering discharge, the peculiarities and difficulties involved with a representative description of these complicated many-particle systems are worked out. Specifically, the scenario is assessed initially from the plasma side by means of the particle in cell (PIC) method. From the basic abstraction to specific details, the concept is laid out with a focus on the problem scenario to be tackled. A similar task is exercised for the conceptual design and singular features of a neutral test particle model. Emphasis is put onto the concept of the coupling of both respective models.

Finally, in section~\ref{sec:results} results for the described exemplary discharge are presented with the focus on merely the physical and conceptual plausibility. For the given setup, a number of characteristic features of the plasma as well as the neutral particles are addressed, which may provide comparison and/or guidance for the well-disposed reader. In this context, the two main postulates initially proposed are verified, providing valid justification for more simplified models relying solely on the test particle approach.\cite{trieschmann_transport_2015}

\section*{Acknowledgments}

\noindent This work is supported by the German Research Foundation (DFG) in the frame of Collaborative Research Centre SFB/TRR 87. The authors gratefully thank D.\ Kr\"uger, T.\ Gergs, S.\ Gallian, and R.\ P.\ Brinkmann from the Institute of Theoretical Electrical Engineering, Ruhr University Bochum and S.\ Ries, N.\ Bibinov, and P.\ Awakowicz from the Institute for Electrical Engineering and Plasma Technology, Ruhr University Bochum for fruitful discussions.

\bibliography{references}
\bibliographystyle{ieeetr}

\end{document}